\documentstyle[psfig]{mn}

\newif\ifAMStwofonts
\newcommand{\be}{\be}

\ifoldfss
  \ifCUPmtlplainloaded \else
    \NewTextAlphabet{textbfit} {cmbxti10} {}
    \NewTextAlphabet{textbfss} {cmssbx10} {}
    \NewMathAlphabet{mathbfit} {cmbxti10} {} % for math mode
    \NewMathAlphabet{mathbfss} {cmssbx10} {} %  "   "    "
  \fi
  \ifAMStwofonts
    \ifCUPmtlplainloaded \else
      \NewSymbolFont{upmath} {eurm10}
      \NewSymbolFont{AMSa} {msam10}
      \NewMathSymbol{\upi}     {0}{upmath}{19}
      \NewMathSymbol{\umu}     {0}{upmath}{16}
      \NewMathSymbol{\upartial}{0}{upmath}{40}
      \NewMathSymbol{\leqslant}{3}{AMSa}{36}
      \NewMathSymbol{\geqslant}{3}{AMSa}{3E}

       \let\le=\leqslant
       \let\ge=\geqslant
    \fi
  \fi
\fi % End of OFSS

\ifnfssone
  \newmathalphabet{\mathit}
  \addtoversion{normal}{\mathit}{cmr}{m}{it}
  \addtoversion{bold}{\mathit}{cmr}{bx}{it}
  \newmathalphabet{\mathbfit} % math mode version of \textbfit{..}
  \addtoversion{normal}{\mathbfit}{cmr}{bx}{it}
  \addtoversion{bold}{\mathbfit}{cmr}{bx}{it}
  \newmathalphabet{\mathbfss} % math mode version of \textbfss{..}
  \addtoversion{normal}{\mathbfss}{cmss}{bx}{n}
  \addtoversion{bold}{\mathbfss}{cmss}{bx}{n}
  \ifAMStwofonts
    \ifCUPmtlplainloaded \else
      \UseAMStwoboldmath
      \makeatletter
      \new@mathgroup\upmath@group
      \define@mathgroup\mv@normal\upmath@group{eur}{m}{n}
      \define@mathgroup\mv@bold\upmath@group{eur}{b}{n}
      \edef\UPM{\hexnumber\upmath@group}
      \new@mathgroup\amsa@group
      \define@mathgroup\mv@normal\amsa@group{msa}{m}{n}
      \define@mathgroup\mv@bold\amsa@group{msa}{m}{n}
      \edef\AMSa{\hexnumber\amsa@group}
      \makeatother
      \mathchardef\upi="0\UPM19
      \mathchardef\umu="0\UPM16
      \mathchardef\upartial="0\UPM40
      \mathchardef\leqslant="3\AMSa36
      \mathchardef\geqslant="3\AMSa3E

       \let\le=\leqslant
       \let\ge=\geqslant
    \fi
  \fi
\fi % End of NFSS release 1

\ifnfsstwo
  \DeclareMathAlphabet{\mathbfit}{OT1}{cmr}{bx}{it}
  \SetMathAlphabet\mathbfit{bold}{OT1}{cmr}{bx}{it}
  \DeclareMathAlphabet{\mathbfss}{OT1}{cmss}{bx}{n}
  \SetMathAlphabet\mathbfss{bold}{OT1}{cmss}{bx}{n}
  \ifAMStwofonts
    \ifCUPmtlplainloaded \else
      \DeclareSymbolFont{UPM}{U}{eur}{m}{n}
      \SetSymbolFont{UPM}{bold}{U}{eur}{b}{n}
      \DeclareSymbolFont{AMSa}{U}{msa}{m}{n}
      \DeclareMathSymbol{\upi}{0}{UPM}{"19}
      \DeclareMathSymbol{\umu}{0}{UPM}{"16}
      \DeclareMathSymbol{\upartial}{0}{UPM}{"40}
      \DeclareMathSymbol{\leqslant}{3}{AMSa}{"36}
      \DeclareMathSymbol{\geqslant}{3}{AMSa}{"3E}

       \let\le=\leqslant
       \let\ge=\geqslant
    \fi
  \fi
\fi % End of NFSS release 2

\ifCUPmtlplainloaded \else
  \ifAMStwofonts \else % If no AMS fonts
    \def\upi{\pi}
    \def\umu{\mu}
    \def\upartial{\partial}
  \fi
\fi

\title[Pistinner \& Eichler:  Self Inhibiting Heat-Flux]
{Self Inhibiting Heat-Flux \\ I. Whistlers Quasilinear Theory}
\author[S.L. Pistinner \& D. Eichler]
  {S.L. Pistinner$^{1}$ \& D. Eichler$^{2}$\\ 
  $^{1}$Department of Applied Mathematics \\
       Israel Institute for Biological Research \\ 
       P.O.B 19 Nes-Ziona 74100 Israel \\
shlomi@math.iibr.gov.il \\
$^{2}$Department of  Physics, Ben-Gurion University \\
Beer-Sheba, 84105, Israel \\
eichler@bguvms.bgu.ac.il}
\date{Accepted xx. Received xx}

\pubyear{1995}

\begin{document}

\maketitle

%\label{firstpage}

\begin{abstract}
Heat-transfer through weakly magnetized diffuse 
astrophysical plasmas 
excites whistlers. This leads to electron whistler
resonant scattering, a reduction of the electron mean-free
path, and heat-flux inhibition.
However, only whistlers propagating at a finite angle to the 
magnetic field
(off-axis) can scatter the heat-flux carrying electrons. Thus,  
the level of heat flux-inhibition along the magnetic field
lines depends on the presence of off-axis whistlers.

We  obtain a solution of the Boltzmann equation with the whistler
wave equation and show that if $\epsilon^{th}\beta_{e} \gg 10^{-4}$,
where $\epsilon^{th}$ is the thermal Knudsen number,
and $\beta_{e}$ is the ratio of the electron pressure to the
magnetic energy density, scattering of heat-flux carrying electrons by
off-axis whistlers,  which are shown to propagate at about $65^{o}$, 
is efficient enough to lead to 
heat-flux inhibition along field lines.
The inhibition so obtained  is 
proportional to $(\epsilon^{th} \beta_{e})^{-1}$. 
\end{abstract}


\section{INTRODUCTION}
The presence of a cold gas embedded in a hot gas is 
a  common phenomenon. 
This situation occurs in many 
astrophysical systems  such as the interstellar 
medium (ISM), the intracluster gas in 
galaxy clusters (ICM),  AGNs and  Lyman $\alpha$ systems.
These multi temperature-density
configurations were and still are the subject of intensive
theoretical study. 
Field (1965) carried out a linear stability analysis of optically 
thin, collision dominated  magnetized plasma. 
He found the critical wavelength (the Field length)
that distinguishes perturbation wavelengths that form cold 
clumps from those that don't. 
Perturbation wavelengths shorter than the Field length
are heated by heat-conduction and fail to form cold clumps  
whereas the longer perturbation wavelengths 
are cooled  enough by radiation losses 
to condense into cold clumps.  The Field length 
retains its qualitative 
role in the non-linear regime (Cowie \& Songaila 1977; 
Cowie \& McKee 1977; McKee \& Cowie 1977).

At large Knudsen numbers there is  heat-flux saturation.   
This is considered by
Cowie \& McKee (1977),
Balbus \& McKee (1982), Slavin \& Cox (1992), and
Dalton \& Balbus (1993).   
Chun \& Rosner (1993)  consider
the effects of a non-local 
Maxwellian electron distribution on Field's analysis, and  
Bandiera \& Chan (1994a,b) generalize their work and 
emphasize the role
of the non-local thermoelectric effect in thermal evaporation 
processes.
Apart from  the details considered in these works, the Field length
retains its qualitative and quantitative role. All the collision dominated 
plasma theories predict that 
scales smaller than the Field length evaporate (heated). 
For typical ISM or ICM parameters,  
the thermal evaporation theory predicts a hotter 
ISM and  ICM  than what is actually 
observed.
This is also the case for Lyman-$\alpha$ systems and the 
standard 
AGN model. 

For the ISM,  McKee \& Ostriker (1977) proposed that 
the observed {\it cold} gas is constantly 
replenished by supernovae explosions.
However, this does not necessarily apply to other 
astrophysical environments
in which cold and hot gas are observed to co-exist.
In  broad line emission regions surrounding quasistellar 
objects (Begeleman 
\& McKee 1990; McKee \& Begeleman 1990) cold and hot gas 
coexist, but the 
presence of a small scale replenishment source is not clear.
The ICM at the cores of some clusters is composed 
of interacting cold and hot gas 
(Canizares, Market \& Donahue 1988; Fukazawa et al. 1994).
Detailed studies show that thermal evaporation as
predicted by classical theory (Spitzer 1962)
should be very efficient in the ICM 
( Cowie \& Binney 1977;  Binney \& Cowie  1981;
Fabian \& Nulsen 1977; Loewenstein \& Fabian 1990; Fabian 
1994).
These works interpret the ICM 
observations as 
a reduction of the magnitude of the thermal conductivity 
below that predicted by classical theory.   
The evidence for heat 
flux inhibition at the ISM, ICM, AGNs, and Lyman-$\alpha$ 
systems is indirect. 
Direct conformation of heat flux inhibition exists in the solar
 corona 
and wind for three decades (Montgomery, Bame \& 
Hundhausen 1968). Recent measurements performed with 
Ulysses (Scime et al. 1994) confirm that between 1.2 and 5.4 
AU, heat flux 
is inhibited compared to its Spitzer value.

This work concentrates on the ICM observations, with a strong 
emphasis on the "cooling flow model" (CFM;
 cf. Fabian et al. 1984, 1991 Fabian 1994 for a review).
The CFM reproduces the observed properties of the 
ICM in the central part of galaxy clusters very well, 
but only with the  assumption
of strong heat flux inhibition, i.e. heat flux is 
ignored. The justification of this hypothesis has been 
heuristic, and has presumed that heat flux inhibition
is obtained by a tangled magnetic field ("in a manner which is 
not yet clear").
This paper considers heat flux inhibition in the 
frame work of a rigorous discussion, and attempts to quantify 
the 
level of heat flux inhibition.
In doing so we partially rely on  works that
consider heat flux inhibition in the solar wind 
( Gary \& Feldman 1977; Gary et al. 1994 and Scime et al. 
1994), due to whistlers.

The ICM observations  require  heat conductivity suppression 
(heat flux inhibition)  compared to the Spitzer value by 
at least two orders of magnitude, and sometimes up to  four
orders of magnitude (Pistinner \& Shaviv 1996; Pistinner,
Levinson \& Eichler 1996).  

Reducing the electron mean free path would lead to heat flux inhibition.
Mechanisms that would accomplish this can be divided 
into two main theoretical classes: 
\begin{enumerate}
\item Increase of the path length of the magnetic field lines,
hence the length over which parallel heat transport must 
operate.
\item Plasma waves that resonantly scatter the heat conducting
 electrons.
\end{enumerate}
The first mechanism is realized when a tangled 
magnetic field pervades the plasma 
(for details cf. Cowie \& Binney 1981, 
Rosner \& Tucker 1989; Tao 1995; Pistinner \& Shaviv 1996). 
The second comes about when self excited electron plasma 
"waves exist" (Gary \& Feldman 1977; 
Jafelice 1992; Levinson \& Eichler 1992; Pistinner et al. 1996).
These plasma waves are expected to scatter 
the heat carrying electrons, thereby reducing their mean free 
path.  

The classification made above is quantitative 
(Goldstein, Klimas \& Sandri 1975), and it follows from 
the different physical processes involved.
More specifically, let $l_{B}$ and $r_{l}$ be the coherence 
length of the 
magnetic field and the electron gyroradius respectively, then: 
\begin{enumerate}
\item Mechanism (i) (the tangling mechanism) is realized when 
the condition
$r_{l}\ll l_{B}$  holds, and does not scatter electrons 
(except for adiabatic mirroring).
\item Mechanism (ii), in which  electromagnetic plasma waves 
scatter electrons
``non-adiabatically'' works if  $r_{L}\approx l_{B}$. 
\end{enumerate}

Large scale tangling, (i), is now seriously constrained and in any case
"questionable" (Tao 1995; 
Pistinner \& Shaviv 1996).
It seems that only a very particular family of magnetic field
correlation functions can inhibit heat conduction.  
These correlation functions have Lorentzian rather than 
Gaussian properties, and the physical basis for their origin is not yet
clear. 
In contrast electromagnetic electron plasma waves (ii), 
obtained from a collisionless plasma model (Levinson \& 
Eichler 1992), provide large inhibition 
factors when applied to the CFM (Pistinner, et al. 1996).
The model of Levinson \& Eichler (1992) follows Gary \& 
Feldman (1977) and
invokes whistler-electron interaction to suppress the heat 
transfer.
In light of our findings (Pistinner et al. 1996) 
and the observational information and interpretation accumulated during
the past three
decades, heat conduction theory in diffuse astrophysical  
plasmas should be reexamined.
Scattering by plasma turbulence should be considered along 
with Coulomb collisions. 

Plasma turbulence results from collective phenomena, and 
may
lead to electron-waves resonant pitch-angle scattering. 
Non-magnetized plasmas are studied in detail, and 
under some conditions ion acoustic turbulence leads 
to heat-flux inhibition (Galeev \& Natanzon 1984).
Jaffliche (1992) considers this possibility in the context of the 
CFM.
Although he  ignores strict requirements for ion-acoustic 
waves excitation, he finds insufficient heat inhibition. 
Levinson \& Eichler (1992)  follow Gary \& Feldman (1977) 
and consider heat flux inhibition by whistlers. 
Gary \& Feldman (1977) find that the  inhibition obtained by 
whistler-electron scattering in the (significantly magnetized) solar  
wind is 
not large. Levinson \& Eichler (1992) show that collision dominated 
plasmas are subject to Weibel type instabilities that 
generate
the whistler electromagnetic mode effectively, and find  
strong heat flux inhibition in {\it weakly} magnetized plasmas. The
differences
between 
the conclusions of Gary \& Feldman and 
Levinson \& Eichler (1992) result from different magnitudes of 
the assumed magnetic contribution to total pressures:
Whereas Gary \& Feldman assume comparable magnetic  
and gas pressures as is appropriate for the solar wind, Levinson \&
Eichler (1992) consider a 
weakly magnetized
 plasma.

Levinson \& Eichler (1992)  conclude  that if the magnetic 
pressure is small compared to the gas pressure,
heat-flux inhibition along field lines is effective. In any case, both
treatments neglect geometric details 
 of 
the plasma processes involved.
In particular, the assumption of a BGK scattering operator 
used by
Levinson \& Eichler 1992 (followed by Pistinner et al. 1996) to calculate
 the particle distribution function, 
which neglects
the pitch angle dependence of the scattering,  was not 
explicitly justified.
 The purpose of this paper is to quantitatively assess the 
validity of the BGK  model in view.
 We show that all pitch
angles  scatter at roughly comparable rates, so that
the neglect of the pitch angles dependence (made by
Gary \& Feldman 1977; Levinson \& Eichler 1992, Pistinner et 
al. 1996) 
is quantitatively valid.

The structure of this paper is:  In \S~\ref{sec:gov} we consider the 
governing 
equations for collision dominated and whistler dominated 
plasma.
In \S~\ref{sec:steady} we obtain the formal expression for the 
electron 
distribution function first in the absence of whistlers and then in the
presence of quasilinear situated whistlers.
In \S~\ref{sec:wis} we obtain the critical 
condition for whistlers excitation by a collision dominated 
plasma. 
We then discuss the consistency of the small scale plasma 
turbulence theory, 
and show that off-axis whistlers are required for significant heat flux 
inhibition and that they are 
excited 
by the required amount.  The expression for the 
heat flux vector is considered in \S~\ref{sec:heat}.
We summarize our conclusions in  \S~\ref{sec:con}.

\section{GOVERNING EQUATIONS}\label{sec:gov}
This section outlines the governing equations
that lead to the heat flux.   
The discussion starts with collision dominated 
plasma theory. It concentrates on the assumptions 
made in deriving the inhibited heat flux, 
in particular, the assumption of {\it steady state},
which raises the question of kinetic stability of the plasma. 

Kinetic plasma instabilities can lead to a plasma pervaded by 
stochastic electromagnetic fields.  These fields may scatter 
particles faster 
than binary collisions.
To account for a kinetic instability saturation,
a set of governing equations that describe this phenomena is 
required.

An electron-ion plasma, irrespective of whether 
the plasma is 
collision or collective phenomena dominated, can be described
by two Boltzmann or Fokker-Planck equations (BFPE). 
For sake of  brevity we shall assume that the solution to the 
ion BFPE is given, and consider only the solution to the 
electron BFPE. 
The electron BFPE (e.g. Blandford \& Eichler 1987) reads:
\begin{equation}
\begin{array}{c}
\mu v {\partial f \over \partial l_{\parallel}}
+{e E^{DC}_{\parallel} \over m_{e}}
\left({\partial f \over \partial v}
+ {1 -\mu^{2} \over v} {\partial f \over \partial \mu}\right)
-{v (1-\mu^{2}) \over 2}
{\partial \ln B \over \partial l_{\parallel}} {\partial f \over \partial 
\mu}= \\
{\partial \kern 5pt \over \partial \mu}
{ 1-\mu^{2} \over 2 } \nu {\partial f \over \partial \mu},
\end{array}
\label{eq:drift_kin}
\end{equation}
where $f$, $\mu$, $v$, $B$, $E^{DC}_{\parallel}$, $\nu$, 
$l_{\parallel}$ are
the electrons distribution function,
the cosine of the pitch angle, the electron speed,  the magnetic 
field magnitude,
the self consistent DC electric field
\footnote{Note that Blandford \& Eichler (1987) do not 
consider the self consistent DC electric field
explicitly. In the case of cosmic rays the relative 
velocity of electrons compared to that of the ions  
is small in comparison to the velocity of the plasma. }
the electron scattering rate (collision frequency) 
\footnote{Note that the scattering rate of the
 electrons can be due to collisions 
with ion or collective phenomena (or both of them and it has 
not yet been 
specified).}, and the affine length
along the magnetic field lines respectively.

Several assumptions have been made in writing 
eq.~\ref{eq:drift_kin}:
\begin{enumerate}
\item steady state
\item Lorentzian plasma, i.e. energy exchange between particles is ignored
\item curvature drifts are ignored
\item phase velocity of the whister waves is ignored.
\end{enumerate}
These assumptions are justified in the context of clusters of galaxies,
where the large scale hydrodynamical timescale and 
the collisional timescale are both long 
compared to  the plasma timescales, the scale length of the magnetic 
field 
is enormous compared to particle gyroradii, and the whistler 
phase velocity is small compared to the electron thermal velocity.
For completeness we elaborate briefly on each of  these
assumptions  in \S~\ref{sec:first_appe}.

Classical astrophysical plasmas  \footnote{With the exception of cosmic 
ray theory.} have traditionally been assumed
to be thermal i.e. collision dominated. 
Thus, the  Spitzer \& H\"arm (1953)  results 
are traditionally applied to model them. With eq.~\ref{eq:drift_kin}
one readily reproduces the Spitzer \& H\"arm (1953) results 
(within a factor of five; cf. Table III and eq. 36 in Spitzer \& H\"arm
as energy exchange is ignored in eq. 1)
along a single magnetic field line. 
Toward that goal one only need  assume that 
Coulomb forces (collision dominated plasma) yield:
\begin{equation}
\nu=\nu_{coll},
\label{eq:coll_coll}
\end{equation}
where
\begin{equation}
\tilde{\nu} \equiv \frac{\nu}{\nu^{th}} =\tilde{\nu}_{coll}=
\tilde{v}^{-3},
\label{eq:coll_coll1}
\end{equation}
and
\begin{equation}
\begin{array}{c}
\nu_{th}=\nu_{coll}(v_{th})=\frac{\omega_p} {N_{D}} \ln 
(\Lambda_{coul})\\
\tilde{v}=\frac{v}{v_{th}}~~v_{th}=\sqrt{\frac{2 k_B T}{m_e}},
\end{array}
\label{eq:coll_coll2}
\end{equation}
with $T$, $k_{B}$, $m_e$, $\omega_{p}$, $N_{D}$ and $\ln 
(\Lambda_{coul})$
the gas temperature, the
Boltzmann constant,  the electron mass, the electron plasma 
frequency, the 
Debye number, and the Coulomb logarithm respectively .

To obtain significant electron heat-flux inhibition 
by whistlers, whistler-electron scattering must be as fast as
ion-electron scattering. 
Thus, the time scale for establishing steady state on 
microscopic 
scales is at least as fast as that when whistlers are absent, 
and we may assume that a modified steady state is formed. 
Under such a situation eq.~\ref{eq:drift_kin} describes the 
evolution 
of the electron distribution function but, with the following 
important modification,
\begin{equation}
\tilde{\nu}=\tilde{v}^{-3}+\tilde{\nu}_{w},
\label{eq:tile}
\end{equation}
where \footnote{The details of the procedure that lead to 
the modifications  to eq.~\ref{eq:drift_kin} in which 
eq.~\ref{eq:tile} 
replaces eq.~\ref{eq:coll_coll} are given by Melrose (1980). }
 $\tilde{\nu_{w}}=\nu_{w}/\nu_{th} $, and $\nu_{w}$ is the 
electron whistler
scattering rate.
A few more implicit assumptions are made in writing  
eq.~\ref{eq:drift_kin}:
The equation is written in a frame in which the whistlers` 
electric field vanishes. 
We assume that the magnitude of the DC field is not big 
enough to lead to 
inelastic scattering events.  The frame in which the scattering 
is elastic moves
with the whistler phase velocity, which has been
neglected in comparison to the velocity of the electrons
exciting the whistlers.

To obtain $\nu^{w}$ one has to choose an approximated non-
linear 
plasma turbulence theory. In this paper we use {\it quasilinear 
theory}. There are several formulation of quasilinear theory, 
and we chose to use the semi-classical formulation given by 
Melrose (1980). From 
his work, after some algebra, we obtain 
\footnote{Melrose 1980; VII, p. 20, f. 7.56; non-relativistic 
electrons assumed}: 
\begin{equation}
\begin{array}{c}
\nu_{w}= \Theta (\mu)\nu_{w,s}+[1-
\Theta(\mu)](\nu_{w,s}+\nu_{w,a})=\\
\nu_{w,s}+[1-\Theta(\mu)](\nu_{w,a}).
\end{array}
\label{eq:tot_scatt}
\end{equation}
where
\begin{equation}
\Theta(\mu)=\Biggl\lbrace\begin{array}{cc}
0 & \mu<0 \\
1 & \mu>0 \end{array},
\end{equation}
\begin{equation}
\begin{array}{c}
\nu_{w,s}={1 \over 2} { \pi^{2} e^2 \over m_{e} c^{2} p \vert \mu 
\vert}
\int_{-1}^{1} { W(k_{r},\chi)\over \chi^{2}} (1-\chi)^{2} d \chi \\
\nu_{w,a}= 2 { \pi^{2} e^2 \over m_{e} c^{2} p \vert \mu \vert}
\int_{-1}^{1} {W(k_{r},\chi)\over \chi}  d \chi \\
k_{r}={\Omega_{e} \over \vert \mu \vert \vert \chi \vert v},
\end{array}
\label{eq:wis_scatt}
\end{equation}
and  $W(k,\chi)$, $\Omega_{e}$, $k$, $e$, $c$, $\chi$, $p$, 
are the
whistler spectrum, the electron gyro-frequency, the
whistler wave number, the electron charge, the speed of light, 
the cosine of the angle
between the whistler propagation direction to the field line 
direction,
and the electron momentum respectively.
Although the above expressions can be written more compactly, 
(Melrose 1980), we  
write them in this way to reflect that $\nu_{w,a}$, has 
contribution from 
asymmetric wave emission while $\nu_{w,s}$ has contribution 
from  symmetric wave emission.
\footnote{The goal of this paper 
is to establish that $\tilde{\nu}_{w,s}\gg 1$ which is the implicit 
assumption made by Levinson \& Eichler. If only  
$\tilde{\nu}_{w,a}\gg1$
then no heat flux inhibition is obtained.}

Equation ~\ref{eq:wis_scatt} demonstrates that 
$\nu$ is a function of  $W(k,\chi)$. 
Thus, one requires an equation that accounts for the evolution 
of the energy density in 
the waves (or the waves spectrum). 
Under the assumption of steady state the equation governing 
the evolution 
of the whistlers spectrum is: 
\begin{equation}
\chi v_{g}\frac{ \partial W}{\partial 
l_{\parallel}}=2W(\gamma^{w}-\Gamma^{w} ),
\label{eq:waves}
\end{equation}
where $\gamma^{w}$ is the quasilinear growth rate (which is 
essentially the linear growth rate
expanded under the assumption $\zeta \equiv \gamma^{w}/ 
\omega^{w}\ll 1$),
$\Gamma^{w}$ is the non-linear damping rate 
\footnote{calculated for an arbitrary mode by second order
perturbation theory (Melrose 1980); and explicitly for whistlers 
by Levinson 1992},
and $v_{g}$ is the whistlers group velocity.
In the framework of quasilinear theory 
the explicit expression for $\gamma^{w}$ (Melrose 1980)
\footnote{Melrose 1980; VII p. 18 f. 7.54}, reads:
\begin{equation}
\begin{array}{c}
\gamma^{w}={\pi^2 k \over 2 n_{e}}
\int_{-1}^{1}d \mu (1-\mu^{2}) 
~~~~~~~~~~~~~~~~~~~~\\~~~~~~~~~
\left\lbrace p^{3} v \left(
\chi {\partial f \over \partial \mu} +
{ m_{i} v_{a}^{2} \over m_{e} v^{2}
\vert \mu \vert} p {\partial f \over \partial p}\right)
\left( 1-\chi {\mu \over \vert \mu \vert}\right)
\right\rbrace_{p=p_{R}},
\end{array}
\label{eq:wisgro0}
\end{equation}
where
\begin{equation}
p_{R}= { m_{e} \Omega_{e} \over k \vert \mu \chi \vert },
\label{eq:resmom}
\end{equation}
is the resonant electron momentum and $m_{i}$ and  $v_{A}$
are the ion mass and the Alfven velocity respectively.  

We have a set of coupled functional equations that govern 
the behavior of the plasma. These equations 
are eq.~\ref{eq:drift_kin} for $f$, 
(with eq.~\ref{eq:tot_scatt}~and~eq.~\ref{eq:wis_scatt} ) and  
eq.~\ref{eq:waves}
for $W$ ( with eq.~\ref{eq:wisgro0}). 
We emphasize that eq.~\ref{eq:drift_kin} and  
eq.~\ref{eq:waves} will 
retain their form in any plasma turbulence theory, and that the
turbulence theory determines eq.~\ref{eq:tot_scatt}, 
eq.~\ref{eq:wis_scatt},
and eq.~\ref{eq:wisgro0}.
With this set of coupled functional equations the distribution 
function 
can be determined and the heat flux can be calculated
self-consistently \footnote{
The functional dependence is implicit:
$f$ is a function of $\nu$ which is a functional of $W$ which is 
a function of $\gamma^{w}$ which is a functional of $f$ etc.}.

\section{SOLUTION OF THE ELECTRONS 
BFPE}\label{sec:steady}
 We consider an
approximate solution to eq.~\ref{eq:drift_kin} 
under the following assumptions:
\begin{enumerate}
\item We assume that a Knudsen expansion is valid, even in 
the
absence of whistler waves.
\item We assume that the net current vanishes.
\end{enumerate}
The first assumption implies that the distribution function 
can be expanded into a series with $f=f^{o}+f^{1}$ and that
$f^{0}\gg f^{1}\gg f^{2}$ such that only the first order 
correction 
can be retained. 

\subsection{The Knudsen Expansion}\label{sec:knud}
The solution obtained by a Knudsen expansion 
can be validated {\it a~posteriori}, and this solution 
procedure is not sensitive to  
whistlers` absence or presence in the plasma.
A little algebra shows that the solution to zeroth order is given by an 
isotropic distribution function  which we take 
to be a Maxwell-Boltzmann distribution. 
With this assumption the first order correction (in 
dimensionless units) reads:
\begin{equation}
{\partial f \over \partial \mu}=
-{\tilde{v} \over \tilde{\nu}}\lambda_{e}^{th}
{\partial f^{MB} \over \partial l_{\parallel}}-
{ e E^{DC}_{\parallel} \over m_e v_{th}\nu_{th} \tilde{\nu}}
{\partial  f^{MB} \over \partial \tilde{v}}.
\label{eq:fir_or}
\end{equation}
where
\begin{equation}
\begin{array}{c}
f^{MB}=n_{e} \pi^{-{3 \over 2}} (m_e v_{th})^{3}e^{-
\tilde{v}^{2}} \\
\epsilon^{coll}= \lambda_{e}^{th}
{\partial \ln (T) \over \partial l_{\parallel}}\tilde{v}^{-4}\equiv
\epsilon^{th} \tilde{v}^{-4},
\end{array}
\label{eq:nut}
\end{equation}
are the Maxwell-Boltzmann distribution,
the Knudsen number, and $\lambda_{e}^{th}=v_{th}/\nu_{th}$ 
is a thermal electron collision mean free path.
Note that only $f^{MB}$ is still dimensional but all other quantities 
are
dimensionless. To eliminate  $E^{DC}$ from 
eq.~\ref{eq:fir_or} we need to relate it to the spatial 
variability of $f^{MB}$.
Toward that goal one uses the current free condition.

\subsection{The Current Free Distribution Function}
In astrophysical conditions, thermoelectric fields would cause the
plasma to settle down
to a zero current state. The current free condition reads:
\begin{equation}
\begin{array}{c}
j_{\parallel}=2 \pi \int d\tilde{v} \tilde{v}^{3}\int d\mu \mu f= \\
2 \pi \int d\tilde{v} \tilde{v}^{3}\int d\mu {1 -\mu^{2}\over 2}
{\partial f \over \partial \mu}\equiv0.
\end{array}
\label{eq:day_kvar}
\end{equation}
Using eq.~\ref{eq:fir_or}, in eq.~\ref{eq:day_kvar} and 
substituting the 
result into  eq.~\ref{eq:fir_or}
we find after some algebra that the current free distribution 
function is: 
\begin{equation}
{\partial f \over \partial \mu}=
-{\tilde{v} \over \tilde{\nu}}
\epsilon^{th} 
\left(\tilde{v}^{2}-\frac{Y^{2}}{Y^{1}}\right)f^{MB},
\label{eq:tcur_free}
\end{equation}
where
\begin{equation}
\begin{array}{c}
Y^{2}=\int_{0}^{\infty} d \tilde{v} \tilde{v}^{6} e^{-\tilde{v}^{2}} 
Z(\tilde{v}) \\
Y^{1}=\int_{0}^{\infty} d \tilde{v} \tilde{v}^{4} e^{-\tilde{v}^{2}} 
Z(\tilde{v}) \\
Z(\tilde{v})\equiv \int_{-1}^{1} d \mu \frac{(1-
\mu^{2})}{\tilde{\nu}},
\end{array}
\label{eq:cur_ints}
\end{equation}
and the ratio of $Y^{2}/Y^{1}$ should be evaluated once the 
value of $\tilde{\nu}$ has been determined. For a collision 
dominated plasma 
the value of $\tilde{\nu}$ is given by  eq.~\ref{eq:coll_coll}. 
Thus, the 
value of   $Y^{2}/Y^{1}$ can be evaluated.
However, if the plasma is whistler dominated, $f$ is  a 
functional  
of $\tilde{\nu}$, and not a simple function of it.
Substituting  $\tilde{\nu}=\tilde{v}^{-3}$ into
eq.~\ref{eq:cur_ints}, one finds $Y^{2}/Y^{1}=4 $ and using this  
in  eq.~\ref{eq:tcur_free},
one obtains
\begin{equation}
\begin{array}{c}
f=f^{MB}\left\lbrack
1+ \mu \epsilon^{coll}
\left( 4- \tilde{v}^{2} \right)\right\rbrack \\
f=f^{MB}\left\lbrack
1+ \mu \epsilon^{th} \tilde{v}^{4}
\left( 4-  \tilde{v}^{2} \right)\right\rbrack.
\end{array}
\label{eq:knud}
\end{equation}
This solution is valid provided that 
$\epsilon^{th} =\epsilon^{coll}(v^{th})
\le 2 \times 10^{-2}$ (due the velocity dependence of the 
collisions mean free
path; Gray \& Killkenny 1980). Note that if 
$\epsilon^{th}=2\times10^{-2}$ then
for $\tilde{v}=2\Rightarrow\epsilon^{coll}=0.32\not\ll 1$, for 
$\tilde{v}=4$
the underlying assumptions of the expansion do not hold.

\section{SOLUTION OF THE QUASILINEAR 
EQUATIONS}\label{sec:wis}
The aim of this section is to a) calculate 
when eq.~\ref{eq:knud} gives rise to whistlers, b) 
to find the angular dependence of the whistler spectrum, and c) 
with this whistler spectrum to estimate the modified steady 
state 
distribution function (to which the collision dominated plasma
distribution function evolves), from which follows the  
heat flux inhibition
factor.  Consistency checks of the {\it a priori} assumptions
made are carried out as we progress. 
To obtain these goals we consider  approximate
solutions to the equations given in \S~\ref{sec:gov}
under the assumption that whistlers pervade the plasma. 
However, before we do so we need to establish when 
this situation would occur and what  the qualitative nature
of it is.

\subsection{The Whistler Quasilinear Growth Rate}
The formal expression for the distribution 
function has been found via the 
Knudsen expansion and is given by
eq.~\ref{eq:tcur_free}. To calculate when whistlers are present 
in the plasma we need to use eq.~\ref{eq:tcur_free} 
\footnote{One can derive the plasma
dispersion relation and consider the 
growth rate for the whistler mode, under the
assumption that the growth rate is much smaller 
then the oscillation frequency
of the waves. This is tantamount to the use of the quasilinear 
growth rate.}
in eq.~\ref{eq:wisgro0}. It proves  
mathematically convenient to formulate all expressions 
in terms of $k_{\parallel}=k\chi$ and $\chi$ as the
independent variables 
rather than $k$ and $\chi$.
After some algebra we obtain the following schematic 
form for the quasilinear growth rate:
\begin{equation}
\gamma^{w}= \frac{\sqrt{\pi}}{2} \Omega_{e}\beta_{e}^{-1} 
\tilde{k}_{\parallel}P(\tilde{k}_{\parallel},\chi)
[\epsilon^{th}\beta_{e}Q(\tilde{k}_{\parallel},\chi) 
P^{-1}(\tilde{k}_{\parallel},\chi)-1]
\label{eq:schem_wis}
\end{equation}
where
\begin{equation}
\begin{array}{c}
P(\tilde{k}_{\parallel},\chi)\equiv 2 \frac{(1+\chi^{2}) }{\chi}
\int_{0}^{1}d \mu (1-\mu^{2}) {1 \over \mu^{5}} 
e^{-\frac{1}{(\tilde{k}_{\parallel} \mu)^2}}= \\
\frac{(1+\chi^{2})}{\chi}e^{-\frac{1}{\tilde{k}_{\parallel}^2}} \\
Q(\tilde{k}_{\parallel},\chi)\equiv \int_{0}^{1} d \mu (1-\mu^{2})
\tilde{v}^{4}e^{-\tilde{v}^{2}} \Biggl\lbrack 
\tilde{v}^{4} \left( \tilde{v}^{2}-\frac{Y^{2}}{Y^{1}}\right)
\\  
\left(\frac{1} {  1+ {\tilde{\nu}^{w,s}\over \tilde{\nu}^{coll}} 
}\right) 
\frac{ (1+\chi^{2}) 
\left(1+ \frac{\tilde{\nu}^{w,s}}{\tilde{\nu}^{coll}}\right) +
(1-\chi)^{2}\frac{\tilde{\nu}^{w,a}}{\tilde{\nu}^{coll}}} 
{1+ \frac{\tilde{\nu}^{w,s}}{\tilde{\nu}^{coll}}+
\frac{\tilde{\nu}^{w,a}}{\tilde{\nu}^{coll}}}
\Biggl\rbrack_{\tilde{v}=\tilde{v}_{r}} \\
\tilde{v}_{r}={1 \over \tilde{k}_{\parallel} \mu}\\
\tilde{k}=k r_{l}^{e} \\
\beta_{e}\equiv\frac{8 \pi P_{e}}{B^{2}}\approx \frac{1}{2} 
\beta\equiv \frac{8 \pi P}{B^{2}}
\end{array}
\label{eq:P_Q}
\end{equation}
and $P_{e}$ is the electrons pressure, and $P$ the total gas 
pressure.
We have chosen the magnetic field pointing from hot to cold.
For this choice whistler propagation is parallel to field lines,
in the coldward direction.  Thus, for
wave growth we must have $\chi>0$ 
(or $\tilde{k}_{\parallel}>0$). This point is taken into account in 
the
expression written above. The terms 
$Q(\tilde{k}_{\parallel},\chi)$ and $P(\tilde{k}_{\parallel},\chi)$  in eq.~\ref{eq:P_Q}
represent a growth term due to the ${\partial {F} \over \partial \mu}$ term,  
and the  damping term due to the ${\partial {F} \over \partial
p}$ term in eq.~\ref{eq:wisgro0}, respectively.

\subsection{ Stability of the Collision Dominated Solution}
Here we show that under a wide range of conditions, 
a distribution function, were it governed only by collisions, would be 
unstable to 
whistler waves which, of course, modify the distribution function.

Using eq. 18 we first establish when a collision 
dominated 
plasma becomes unstable.
Using $Y^{2}/Y^{1}=4 $
and $\tilde{\nu}^{w,s}=\tilde{\nu}^{w,a}=0$ in eq.~\ref{eq:P_Q}, 
we have:
\begin{equation}
\begin{array}{c}
\gamma^{w}_{coll}=\frac{\sqrt{\pi}}{2} \Omega_{e}\beta_{e}^{-
1} (1+\chi^{2}) 
\tilde{k}_{\parallel}
\Biggl\lbrack\int_{0}^{1} d \mu (1-\mu^{2})
\tilde{v}^{4}e^{-\tilde{v}^{2}} \\ 
\Biggl(
\epsilon^{th} \beta_{e} \tilde{v}^{4} \left( \tilde{v}^{2}-4 \right)
\Biggl)_{\tilde{v}=\tilde{v}_{r}}-
\frac{e^{-\frac{1}{(\tilde{k}_{\parallel})^2}}}{\chi}\Biggl\rbrack.    
\end{array}
\label{eq:maxw_gro_coll}
\end{equation}
where the subscript {\it coll} denotes whistler growth 
in a distribution function stabilized by binary collisions only, not
by
the whistlers themselves.
Some rough estimates may be obtained from  
eq.~\ref{eq:maxw_gro_coll}. 
The first is that, provided $\epsilon^{th} \beta_{e}>1$,
we have $\gamma^{w}> 0$, in a narrow cone along the axis. 
The width angle of this cone is roughly estimated to be 
$ \beta_{e}^{-1}(\epsilon^{th})^{-1} < 1$.
To obtain the quantitative conditions we note that
the integral in eq.~\ref{eq:maxw_gro_coll} can be written 
in terms of tabulated functions,
\begin{equation}
\begin{array}{c}
\gamma^{w}_{coll}= 
\sqrt{\pi} \Omega_{e} \beta_{e}^{-1} 
(1+\chi^{2})\tilde{k}_{\parallel}
\Biggl( \frac{\epsilon^{th}\beta_e}{(\tilde{k}_{\parallel})^{8}}
\Biggl\lbrace (\tilde{k}_{\parallel})^{2}
e^{-\frac{1}{(\tilde{k}_{\parallel})^{2}}}+ \\
(\tilde{k}_{\parallel})^{4}e^{-\frac{1}{(\tilde{k}_{\parallel})^{2}}} 
-\frac{15}{8}(\tilde{k}_{\parallel})^{6}
e^{-\frac{1}{(\tilde{k}_{\parallel})^{2}}}-
\frac{9\sqrt{\pi}}{8}(\tilde{k}_{\parallel})^{5} 
erfc[(\tilde{k}_{\parallel})^{-1}] \\ 
+\frac{15\sqrt{\pi}}{16} 
(\tilde{k}_{\parallel})^{7} erfc[(\tilde{k}_{\parallel})^{-1}] 
\Biggl\rbrace- \frac{e^{-
\frac{1}{(\tilde{k}_{\parallel})^{2}}}}{\chi}\Biggl),
\end{array}
\label{eq:matz_a}
\end{equation}
where 
\begin{equation}
erfc(x)=1-\frac{2}{\sqrt{\pi}} \int_{0}^{x} dt e^{-t^2}.
\end{equation}
Marginal stability of the plasma is  found numerically  in
$\beta_{e}\epsilon_{th}, \tilde{k}_{\parallel}, \chi$ space at
\begin{equation}
\begin{array}{c}
 \epsilon_{th}\beta_{e}=0.0001\pm 10^{-5}\Rightarrow \\
\gamma^{w}_{coll}(\tilde{k}_{\parallel}=0.22\pm 10^{-3},\chi=1 
\pm 10^{-4})=0\pm 10^{-11}.
\end{array}
\label{eq:mar_coll}
\end{equation}
The validity of  the latter solution is subject to the condition 
\begin{equation}
\begin{array}{c}
\Omega_{i}\ll\omega^{w} \ll \Omega_{e} \\
0.00054 \ll \tilde{k}^{2} \vert \chi\vert \beta_{e}^{-1}\ll1 \\
0.00054 \ll \frac{\tilde{k}_{\parallel}^{2}}{ \vert \chi\vert} 
\beta_{e}^{-1}\ll1, 
\end{array}
\label{eq:wave_cond}
\end{equation}
where
\begin{equation}
\omega^{w}=\Omega_{e}\tilde{k}^{2}\beta_{e}^{-1} \vert \chi 
\vert,
\end{equation}
which is the range in which the whistler mode exists. 
A numerical study of the properties of 
eq.~\ref{eq:matz_a} shows that 
there is a distinct range of $\tilde{k}_{\parallel}$
for which $\gamma^{w}_{coll}>0$. This value is 
bounded between $0.22<\tilde{k}_{\parallel}<0.7$ for  
$0.0001<\epsilon^{th}\beta_{e}<1000$. 
For  $\tilde{k}_{\parallel}=0.22$, one finds 
\begin{equation}
180\gg\beta\gg0.01.
 \label{eqq:wis_beta_cond_mod}
\end{equation}
For $\tilde
k_{\parallel}=0.7$ instead of $0.22$ we find 
that:
\begin{equation}
1800  \gg \beta \gg 1.
\label{eq:wis_beta_cond_mod}
\end{equation}
 The underlying physical interpretation of this range of $\beta$ is as
follows: If $\beta$ is too large, the waves that resonate with thermal
electrons can cyclotron resonate with ions as well, which complicates
things beyond the scope of this paper. For $\beta$ too small, the phase
velocity of the whistlers exceeds the thermal electron velocity by a
sufficient margin that the resonant particles are far out on the tail of 
the Maxwellian distribution function, and the excitation is too weak
to be interesting. In any case, condition  eq.~\ref{eq:wis_beta_cond_mod} 
is met in clusters of galaxies, and 
condition \ref{eqq:wis_beta_cond_mod} at the ISM. 
The angular dependence of $\gamma$ on
eq.~\ref{eq:matz_a}
 on $\chi$ is roughly linear with a maximum at 
$\chi=1$; the slope  depends on  $\epsilon^{th}\beta_{e}$.
Thus, at $\chi<1$, condition  eq.~\ref{eq:wis_beta_cond_mod}  or condition  eq.~\ref{eqq:wis_beta_cond_mod} are somewhat relaxed by a factor of order
unity.

The critical condition for whistlers present in a 
collision dominated hydrogen plasma (cf. the last line of 
eq.~\ref{eq:P_Q} ) 
 is found from eq.~\ref{eq:mar_coll} to be :
\begin{equation}
\epsilon^{th}\beta > 2\times 10^{-4},
\label{eq:krit_krit}
\end{equation}
provided that $\beta$ is in the range given by 
eqs.~\ref{eq:wis_beta_cond_mod}, ~\ref{eqq:wis_beta_cond_mod}.

There are a few more conditions for the validity of 
eq.~\ref{eq:krit_krit}.
One of them is
\begin{equation}
\epsilon^{th}\le 10^{-2}\Rightarrow \beta > 10^{-2},
\end{equation}
and it stems from the validity of the Knudsen expansion.
The latter constraint is so weak that it can be ignored. 

The last condition to be considered is merely a formal one 
and it stems from the validity of 
eq.~\ref{eq:wisgro0},  namely:
\begin{equation}
\zeta\equiv \frac{\max (\gamma^{w}_{coll}) }{\omega^{w}} \ll 1 
\label{zeta}
\end{equation}
This condition is rather strong and it requires that 
$\epsilon^{th}\beta_{e}<10$.
However, even if this is not the case if when whistlers dominate 
the plasma, 
the value of $\nu_{w,s}$ is sufficiently high and $\zeta$ can be 
maintained much 
smaller than unity.

The conclusion of this section is 
that any hydrodynamical model in which $\beta>0.01$ and
$10^{-4}<\epsilon^{th}<10^{-2}$
gives rise to whistlers due to an excess of perpendicular momentum  in 
the hotward velocity hemisphere.  The unstable on-axis whistlers 
propagate in the coldward direction.

 Note that  the hotward part of the distribution function, which  
has  excess perpendicular momentum is  not the part that 
carries the heat. The coldward particles, which carry 
the heat from hot to cold, 
have an excess of parallel momentum, and are stable to all 
on-axis whistlers

Thus, the heat carrying
particles cannot resonantly
excite on-axis whistlers that propagate parallel 
to the field lines and on-axis   whistlers that propagate parallel to the 
field lines do not significantly inhibit heat flux.
Whistlers 
that propagate off the axis, on the other hand, can inhibit heat flux.
These off-axis waves have right-handed elliptical 
polarity which can be 
represented as a superposition of left-handed and right-handed 
circular polarity. The left-handed circular polarity allows 
for whistler emission in perpendicular momentum deficient 
regions in phase space. 
Although at marginal stability of a collision dominated plasma only 
on-axis whistlers are excited, it is incorrect to proceed  under the assumption 
that these on-axis whistlers 
can lead to heat flux inhibition. 
Rather, we argue, any on-axis waves merely modify the distribution
function in such a way that the marginal stability is transferred to
off-axis waves.

\subsection{Condition for Whistler Dominated 
Plasma and Energy Density of Whistlers}
The questions are related as to when the plasma would become 
whistler dominated and how much energy density would be stored in 
the whistlers,  and we consider 
them together. 
As was pointed out above, a whistler dominated plasma does not 
necessarily imply heat-flux inhibition. 
The plasma becomes whistler dominated when
\begin{equation}
\frac{\tilde{\nu}^{w,a}}{\tilde{\nu}^{coll}}\gg 1,
\label{eq:back}
\end{equation}
whereas the whistler dominance leads to heat flux inhibition 
when
\begin{equation}
\frac{\tilde{\nu}^{w,s}}{\tilde{\nu}^{coll}}\gg 1.
\label{eq:symmm}
\end{equation}

Using  eq.~\ref{eq:coll_coll1} and
the first line of eq.~\ref{eq:wis_scatt} in eq.~\ref{eq:symmm} we 
find that 
heat flux inhibition is obtained provided that:
\begin{equation}
\begin{array}{c}
\frac{\tilde{\nu}^{w,s}}{\tilde{\nu}^{coll}}= \\
\frac{9}{8}\beta_{e}^{-1 \over 2} 
2\frac{N_D}{Log(\Lambda_{culo})}
\left(\frac{v_{th}}{c}\right)\int_{1-\xi}^{1} 
\tilde{W}(\tilde{k}_{R},\chi) (1-\chi^2)\frac{d \chi}{\chi^2} \gg 1.
\end{array}
\end{equation}
Writing the spectrum in the following general way:
\begin{equation}
\tilde{W}=\Sigma_{i} K^{i}(\tilde{k}_{\parallel})\Xi^{i}(\chi),
\label{eq:spec_par}
\end{equation}
where $K^{i}$ are arbitrary functions of $\tilde{k}_{\parallel}$, 
and $\Xi^{i}(\chi)$ are arbitrary functions of $\chi$
we have:
\begin{equation}
\begin{array}{c}
\int_{1-\xi}^{1}
\tilde{W}(\tilde{k}_{R},\chi) (1-\chi^2)\frac{d \chi}{\chi^2} =\\
\Sigma_{i}  
K^{i}(\tilde{v}\mu) 
\int_{1-\xi}^{1} \Xi^{i}(\chi) (1-\chi^2)\frac{d \chi}{\chi^2}.
\end{array}
\end{equation}
Thus, if $\xi$ is sufficiently big  we have heat flux inhibition.
Assuming that this is the case, we write 
$\tilde{W}(\tilde{k}_{R},\chi)=
Const~\tilde{W}(\tilde{k})$ and assume $Const$ is of order 
unity. 
We can now invert the above ratio and obtain the constraint on 
the energy density in the waves which reads: 
\begin{equation}
\begin{array}{c}
{4 \pi W(k_{R})k_{R} \over B^{2}}
\approx \\  \tilde{W}(\tilde{k}) = 
N_D^{-1}\frac{\tilde{\nu}^{w,s}}{\tilde{\nu}^{coll}}
\frac{8}{9}\beta_{e}^{1 \over 2}Log(\Lambda_{culo}) 
\left(\frac{v_{th}}{c}\right)^{-1}.
\end{array}
\end{equation} 
The energy density in the whistlers is a function of the 
inhibition factor, and inversely proportional to it. 
Since the  inverse of the Debye number is tiny,
$10^{-16}$ for the ISM and $10^{-23}$ for the ICM, 
inhibition factors can be extremely large. 
Thus, even  an inhibition of the heat flux
by a factor of $10^{10}$ leaves the energy density in the 
whistlers small
compared to the energy density in the magnetic field.
Therefore, one can quite safely assume that the energy 
density stored
in the whistlers is invariably much less than the energy density 
of the background magnetic field.

\subsection{Whistler Dominated Plasma}
We have shown that once the condition
~\ref{eq:krit_krit} is satisfied the plasma becomes
whistler dominated. We now proceed under the assumption 
that eq. ~\ref{eq:krit_krit}  is satisfied by a large margin, i.e. that
\begin{equation}
\epsilon^{th}\beta \gg  ~ 2\times 10^{-4}.
\end{equation}
Under this conditions whistlers grow as 
$\gamma^{w}_{coll}>0$, for at least the same mode,
and we need to solve the wave equation eq.~\ref{eq:waves}.
Thus, we need to  determine $\tilde{W}$,  and $\tilde{\nu}$  
self-consistently.  We consider  quasilinear saturation 
i.e $\gamma^{w}=0$, for the most unstable mode. In particular we need to
know 
whether quasilinear saturations sets in before or after  
eq.~\ref{eq:symmm} becomes relavant, 
as only the latter leads to heat
flux inhibition. Eq.~\ref{eq:back}, though implying a whistler 
dominated 
plasma, merely implies a reduction of the heat-flux only by a factor
of 
two.

Advection by waves is negligible (cf. \S~ \ref{app:A})
and  the wave equation 
eq.~\ref{eq:waves} with eq.~\ref{eqq:eta_eps} reduces to
\begin{equation}
\gamma^{w}=\Gamma^{w}.
\end{equation}
Quasilinear saturation is defined as the vanishing of $\gamma^{w}$ and
hence of $\Gamma^{w}$. If the wave spectrum achieves quasilinear 
saturation to a good approximation, 
i.e. for every $\tilde{k}_{\parallel}$ there is a critical 
angle $\chi_{crit}$ such that:
\begin{equation}
\gamma^{w}[\tilde{k}_{\parallel},\chi_{crit}(\tilde{k}_{\parallel})
]=0,
\label{eq:quasi_sat}
\end{equation} 
and for all other angles the growth rate $\gamma^{w}$ is 
negative. Thus, 
\begin{equation}
W \propto \delta (\chi-\chi_{crit}).
\label{eqq:quasi_sat}
\end{equation}
The picture is that the wave spectrum is highly peaked 
near $\chi_{crit}$, where the condition of marginal stability is 
met. Once this has been
established, the condition $\zeta \ll 1$ (cf. eq.~\ref{zeta}) is 
{\it a priori} fulfilled.

The assumption of quasilinear saturation is exact if 
$\chi_{crit}$ is unique  and does not depend on 
$\tilde{k}_{\parallel}$. Otherwise  the assumption 
$\Gamma^{w}=0$ is formally invalid
(Mode coupling does not exist if all the waves propagate at the same 
angle). However, the non-linear Landau damping rate 
$\Gamma^{LD}$ is of order 
$\beta_{e}^{-2}$ compared to the mode coupling term, and 
it can be shown (Levinson 1992; cf.~\ref{app:B}) that if 
the parent wave and the resulting waves have energies
of the same order, then the non-linear damping 
 is less than 
$\tilde{\nu}^{w,s}/\tilde{\nu}^{coll} \eta$ ($\eta$ is defined in eq.~\ref{eta} ). 
Thus, it is of order the suppression factor times $\eta$ 
and can quite safely be ignored for the suppression factors 
that we are interested in.

Using eq.~\ref{eq:schem_wis} in eq.~\ref{eq:quasi_sat} we 
obtain:
\begin{equation}
\epsilon^{th}\beta_{e}Q(\tilde{k}_{\parallel},\chi_{crit}) 
P^{-1}(\tilde{k}_{\parallel},\chi_{crit})-1=0
\label{eq:schem_wis_sat}
\end{equation}
We now seek solutions of eq.~\ref{eq:schem_wis_sat} which 
will yield 
a value of $\chi_{crit}$ by exploiting the fact that 
$Q(\tilde{k}_{\parallel},\chi_{crit}) 
P^{-1}(\tilde{k}_{\parallel},\chi_{crit})$ has a single maximum 
of $\chi$ at
$\chi_{crit}$.  To start this procedure we  
 evaluate $Q$, which is defined 
in eq.~\ref{eq:schem_wis}. 
In particular we  note the $Q$ dependence on
the ratios
$\tilde{\nu}^{w,s}/\tilde{\nu}^{coll}$, 
and $\tilde{\nu}^{w,a}/\tilde{\nu}^{coll}$
at $\tilde{v}=\tilde{v}_{r}$. 
Using eq.~\ref{eq:spec_par} in eq.~\ref{eq:wis_scatt}, 
and $\tilde{\nu}^{coll}=\tilde{v}^{-3}$ we find:
\begin{equation}
\begin{array}{c}
\frac{\tilde{\nu}^{w,a}}{\tilde{\nu}^{coll}}\Biggl)_{v=v_r}= \aleph
\left(\frac{\tilde{k}_{\parallel}}{(\tilde{k}_{\parallel} 
\mu)^{3}}\right)
\Sigma_{i} K(\tilde{k}_{\parallel}) {\cal A}_i \\
\frac{\tilde{\nu}^{w,s}}{\tilde{\nu}^{coll}}\Biggl)_{v=v_r}=  
\aleph
\left(\frac{\tilde{k}_{\parallel}}{(\tilde{k}_{\parallel} 
\mu)^{3}}\right)
\Sigma_{i} K(\tilde{k}_{\parallel}) {\cal S}_{i}
\end{array}
\label{eq:coll_ratios}
\end{equation}
where
\begin{equation}
\begin{array}{c}
\aleph \equiv
\left(\frac{3 N_{D} \beta_{e}^{-1 \over 2}}{4 \ln 
(\Lambda_{coul})}
\right)
\sqrt{\frac{m_i}{m_e}} \left(\frac{v_{th}}{c}\right)  \gg 1 \\
{\cal A}_i \equiv 4
\int_{0}^{1} d \chi \frac{\Xi_{i}(\chi)}{\chi} \\
{\cal S}_{i} \equiv
\int_{0}^{1} d \chi
\frac{\Xi_{i}(\chi)}{\chi^{2}}(1-\chi)^{2}.
\end{array}
\label{eq:A_S}
\end{equation}
We consider now  eq.~\ref{eq:symmm} which  
gives the relevant case. 
The elimination of the case 
given by eq.~\ref{eq:back} is considered further below.

A whistler dominated plasma under the condition
eq.~\ref{eq:symmm} implies that
$ \aleph {\cal A}_{i} K(\tilde{k}_{\parallel})
\gg 1$ and $ \aleph {\cal S}_{i} K(\tilde{k}_{\parallel})
\gg 1$ for the dominant $i$. Substituting 
eq.~\ref{eq:coll_ratios} into eq.\ref{eq:P_Q} we find
\begin{equation}
\begin{array}{c}
Q(\tilde{k}_{\parallel},\chi)= \int_{0}^{1} d \mu (1-\mu^{2})
\frac{\tilde{k}_{\parallel}}{(\tilde{k}_{\parallel}\mu)^{8}}
e^{- \frac{1}{(\tilde{k}_{\parallel}\mu)^{2}}} 
\left( \frac{1}{(\tilde{k}_{\parallel}\mu)^{2}}-\frac{Y^{2}}{Y^{1}}\right)\\
\frac{(\tilde{k}_{\parallel}\mu)^{3}\tilde{k}_{\parallel} }
{\Sigma_{i} K_{i}(\tilde{k}_{\parallel}) ({\cal A}_{i}+{\cal S}_{i})}
\Biggl\lbrack 
(1+\chi^{2}) + (1-\chi)^{2} 
\frac{\Sigma_{i} K_{i}(\tilde{k}_{\parallel}) {\cal A}_{i}}
{\Sigma_{i} K_{i}(\tilde{k}_{\parallel}) {\cal S}_{i}}
\Biggl\rbrack.
\end{array}
\end{equation}
Using the last equation in 
eq.~\ref{eq:schem_wis_sat}
we obtain:
\begin{equation}
\begin{array}{c}
\epsilon^{th}\beta_{e}
\frac{G^{5}\tilde{k}_{\parallel}^{4}}
{\aleph \Sigma_{i} K(\tilde{k}_{\parallel})({\cal S}_{i}+{\cal 
A}_{i})}
 \\
\left(\chi_{crit}+\frac{\Sigma_{i} K(\tilde{k}_{\parallel}){\cal A}_{i}}
{\Sigma_{i} K(\tilde{k}_{\parallel}) {\cal S}_{i}}
\frac{ \chi_{crit} (1-\chi_{crit})^{2}}{1+\chi_{crit}^{2}}\right)\approx 1 
\end{array}
\label{eq:off_waves}
\end{equation}
where
\begin{equation}
G^{n}(\tilde{k}_{\parallel})\equiv 
\frac{1}{\tilde{k}_{\parallel}^{8}}
e^{\frac{1}{\tilde{k}_{\parallel}^{2}}}
\int_{0}^{1} \frac{ d \mu (1-\mu^{2})}{\mu^{n}}
\left(\frac{1}{(\tilde{k}_{\parallel} \mu)^{2}}-
\frac{Y^{2}}{Y^{1}}\right)
e^{\frac{-1}{(\tilde{k}_{\parallel}\mu)^{2}}},
\end{equation}
is a rational function of $\tilde{k}_{\parallel}$, and 
\begin{equation}
G^{5}=\frac{1}{2}\tilde{k}_{\parallel}^{-8} 
\left[\tilde{k}_{\parallel}^{2}+\tilde{k}_{\parallel}^{4}
\left(2-\frac{Y^2}{Y^1}\right)\right].
\label{eq:G5}
\end{equation}
Substituting eq.~\ref{eq:G5} into eq.~\ref{eq:off_waves}
one finds:
\begin{equation}
\begin{array}{c}
\epsilon^{th}\beta_{e}
\left(\frac{1}{\tilde{k}_{\parallel}^{2}}+2-\frac{Y^2}{Y^1}\right)
\\
~~~~~\frac{1}{ 2 \aleph \Sigma_{i} 
K(\tilde{k}_{\parallel})({\cal S}_{i}+{\cal A}_{i})}
\left(\chi_{crit}+\frac{\Sigma_{i} K(\tilde{k}_{\parallel}){\cal A}_{i}}
{\Sigma_{i} K(\tilde{k}_{\parallel}) {\cal S}_{i}}
\frac{ \chi_{crit} (1-\chi_{crit})^{2}}{1+\chi^{2}_{crit}}\right)\approx 1,
\end{array}
\end{equation}
 If we  proceed under the assumption that 
only 
a single value of $i$ contributes to the sum, we find:
\begin{equation}
\begin{array}{c}
\epsilon^{th}\beta_{e}
 \left(\frac{1}{\tilde{k}_{\parallel}^{2}}+2-
\frac{Y^2}{Y^1}\right)\\~~~~~
\frac{1} {2 \aleph  K(\tilde{k}_{\parallel}){\cal S}(1 +{\cal R})}
\left(\chi_{crit}+{\cal R} \frac{ \chi_{crit} (1-
\chi_{crit})^{2}}{1+\chi^{2}_{crit}}\right)\approx 1,
\end{array}
\label{eq:R_S}
\end{equation}
where
\begin{equation}
{\cal R}=\frac{\cal A}{\cal S}.
\end{equation}
We can now obtain the critical angle for whistlers propagation 
at  marginal quasilinear stability. Note that the function given 
in the parenthesis might have a local maximum in $\chi_{crit}$. 
One finds this maximum from the condition:
\begin{equation}
\begin{array}{c}
\frac{1} {\cal R}+\frac{ (1-
\chi_{crit})^{2}}{1+\chi^{2}_{crit}}+\\ \frac{ \chi_{crit} (1-4
\chi_{crit}+3\chi_{crit}^{2})}{1+\chi^{2}_{crit}}
\frac{ 2 \chi_{crit}^{2} (1-
\chi_{crit})^{2}}{(1+\chi^{2}_{crit})^{2}}=0
\end{array}
\label{eqq:max_con}.
\end{equation}
Thus, the  location of the maximum for the value of 
$\chi_{crit}$ depends on the value 
${\cal R}$ only.
Using eq.~\ref{eqq:quasi_sat} in eq.~\ref{eq:A_S} to 
estimate the value of ${\cal R}$ we find:
\begin{equation}
{\cal R} \approx \frac{ 4 \chi_{crit}}{(1-\chi_{crit})^{2}}.
\label{eq:RRR}
\end{equation}
 The cubic 
 equation eq.~\ref{eqq:max_con} 
has two complex roots and one real root.
Substituting into the expression for the real root the 
value of ${\cal R}$ from eq.~\ref{eq:RRR}, 
we find an equation for $\chi_{crit}$.
We solve this equation numerically, 
and find that
\begin{equation}
{\cal R}=4.8,
\end{equation}
and
\begin{equation}
\chi_{crit}=0.414214 \Rightarrow \theta\approx 64^{o}.
\end{equation}
We now substitute the values  of $\chi_{crit}$ and ${\cal R}$
into eq.~\ref{eq:R_S} and find:
\begin{equation}
\epsilon^{th}\beta_{e}
 \left(\frac{1}{\tilde{k}_{\parallel}^{2}}+2-\frac{Y^2}{Y^1}\right)
\frac{5.88} {2 \aleph  K(\tilde{k}_{\parallel}) {\cal S}} \approx 1.
\end{equation}
One can now solve this equation for the value of 
$\aleph  K(\tilde{k}_{\parallel}) {\cal S}$. 
Bearing in mind that only one $i$ contributes to the 
sum in eq.~\ref{eq:coll_ratios} and substituting 
the value of $\aleph  K(\tilde{k}_{\parallel}) {\cal S}$
into the last line of eq.~\ref{eq:coll_ratios} we find:
\begin{equation}
\tilde{\nu}^{w,s}\approx
\frac{3 \epsilon^{th}\beta_{e}} {\mu \tilde{v}}\left[
(\tilde{v} \mu)^{2} +2-
\frac{Y^1}{Y^2}\right],
\label{eqq:sol_s}
\end{equation}
and with the numerical value of ${\cal R}$ 
we have  from the first line of eq.~\ref{eq:coll_ratios}
\begin{equation}
\tilde{\nu}^{w,a}\approx
\frac{14 \epsilon^{th}\beta_{e}}{\tilde{v}{\mu}} \left[
(\tilde{v} \mu)^{2} +2-
\frac{Y^2}{Y^1}\right],
\label{eqq:sol_a}
\end{equation}
which  implies that the heat flux along field lines
is roughly proportional to $\beta_{e}^{-1}$, and 
can be determined from observable parameters. 
The solution procedure continues by finding the value 
of the ratio $Y^{2}/Y^{1}$. This problem  
can be considered separately, and we do so in \S~\ref{app:B}. 
Thus, with eqs.~\ref{eqq:sol_a},~\ref{eqq:sol_s} 
and eq.~\ref{eq:tcur_free} we have:
\begin{equation}
\begin{array}{c}
{\partial f \over \partial \mu}\approx
-{\tilde{v}^2 \mu \over 3 +14 [1-\Theta(\mu)]} \\
\beta_{e}^{-1}
 \left(\tilde{v}^{2}-\frac{Y^{2}}{Y^{1}}\right)
\left[ (\mu \tilde{v})^{2}+2-\frac{Y^{1}}{Y^{2}}\right]^{-1}f^{MB},
\end{array}
\label{eqq:tcur_free}
\end{equation}
and from \S~\ref{app:B} we find  
$Y^{2}/Y^{1}\approx 0.18$. Thus,  the distribution function reads:
\begin{equation}
\begin{array}{c}
{\partial f \over \partial \mu}\approx
-{\tilde{v}^2 \mu \over 3 +14 [1-\Theta(\mu)]} \\
\beta_{e}^{-1}
 \left(\tilde{v}^{2}-0.18\right)
\left[ (\mu \tilde{v})^{2}+1.82\right]^{-1}f^{MB},
\end{array}
\label{eqqq:tcur_free}
\end{equation}
and a self consistent distribution function  at quasilinear saturation 
of off-axis waves has been found.

Although in a collision dominated  plasma whistlers are least stable 
 along the axis we now show  that
at  quasilinear (QL) saturation this is no longer the case:  
If the waves at QL saturation would have 
propagated in a narrow cone along the axis this would have implied 
that   $ \aleph {\cal A}_{i} K_i(\tilde{k}_{\parallel}) 
\gg 1$ for some $i$ and that $
 {\cal S}_{i}\ll {\cal A}_{i} $  for every $i$.
Substituting this assumption into eq.~\ref{eq:schem_wis_sat} 
we find:
\begin{equation}
\epsilon^{th}\beta_{e} \left( \chi_{crit} \frac{ 
G^{5}(\tilde{k}_{\parallel}) 
\tilde{k}_{\parallel}^{4}}
{\aleph \Sigma_i K(\tilde{k}_{\parallel}){\cal A}_{i}}
+G^{8} (\tilde{k}_{\parallel}) \frac{\chi_{crit} (1-\chi_{crit})^{2}}
{1+\chi_{crit}^{2}}\right)\approx 1
\label{eq:axis_growth}
\end{equation}
Now $G^{5}$ and $G^{8}$ are functions of the same order of 
magnitude.  
The function in the parenthesis  has a local maximum at $\chi_{crit}$.
Provided we neglect the term of order $1/\aleph$ the 
location of the maximum does not
depend on $\tilde{k_{\parallel}}$ or  
 $\epsilon^{th}\beta$ and it is found at $\chi_{crit}\approx 0.3$,
in contradiction with our assumption that $\chi_{crit}\approx 1$
that ${\cal S}_{i} $ is of order  ${\cal A}_{i} $.
Since the corrections to the location of the maximum  would only be 
of order $1/\aleph $, we conclude that it is very difficult to support 
a whistler dominated plasma that does not generate off-axis
waves without imposing strict fine tuning.

\section{THE HEAT-FLUX ALONG FIELD 
LINES}\label{sec:heat}
The heat flux vector is, after integration by parts,
\begin{equation}
q_{\parallel} =- \pi m_{e} \int_{0}^{\infty} dv v^{5} \int_{-1}^{1}
d\mu (1-\mu^{2}){\partial f \over \partial \mu}.
\label{eq:heat_fon}
\end{equation}
When eq.~\ref{eqqq:tcur_free} is used in 
eq.~\ref{eq:heat_fon} one gets:
\begin{equation}
\begin{array}{c}
q_{\parallel} =-\pi^{-{1\over 2}} m_{e} v_{e}^{3} n_{e} 
\epsilon^{th}  (\beta_{e} \epsilon^{th})^{-1} \\
\int_{0}^{\infty} 
d \tilde{v} \tilde{v}^{7} e^{-\tilde{v}^2}
\left(\tilde{v}^{2}-0.18\right)
\int_{0}^{1} d \mu (1-\mu^{2})  \mu \\
{1 \over 3 +14 [1-\Theta(\mu)]}
\beta_{e}^{-1}
\left[ (\mu \tilde{v})^{2}+1.82\right]^{-1}.
\end{array}
\label{eq:heat_tran}
\end{equation}
Carrying the integration over the angles yields
\begin{equation}
\begin{array}{c}
q_{\parallel} =\frac{14}{51} 
\frac{\pi^{-{1\over 2}} }{2} m_{e} v_{e}^{3} n_{e} 
\epsilon^{th}  (\beta_{e} \epsilon^{th})^{-1} \\
\int_{0}^{\infty} 
d \tilde{v} \tilde{v}^{3} e^{-\tilde{v}^2}
\Biggl[1.82 \left(1+\frac{\tilde{v}^{2}}{1.82}\right)\log
\left(1+ \frac{\tilde{v}^{2}}{1.82}\right)- \tilde{v}^{2}\Biggl].
\end{array}
\end{equation}
Carrying the integral over the velocities we find:
\begin{equation}
\begin{array}{c}
q_{\parallel} =0.13 \pi^{-{1\over 2}} m_{e} v_{e}^{3} n_{e} 
\epsilon^{th}  (\beta_{e} \epsilon^{th})^{-1} \\
\Biggl[\Gamma(3.5)-0.18\Gamma(2.5)+0.083(
-2(1.82)+(1.82)^2+1.82^{3}- \\
6 e^{1.82} G^{4,0}_{4,0} \left(1.82 \Biggl\vert \begin{array}{c} \phi,0,0 \\ -1,-1,3,\phi\end{array}\right) + \\
4 e^{1.82} G^{4,0}_{4,0} \left(1.82 \Biggl\vert \begin{array}{c} \phi,1,1 \\ 0
,0,4,\phi\end{array}\right)-  \\
e^{1.82} G^{4,0}_{4,0} \left(1.82 \Biggl\vert \begin{array}{c} \phi,2,2 \\ 1,1,5,
\phi\end{array}\right) \Biggl]\\
+0.25*0.18\Biggl[1.82+(1.82)^2+2  \\
e^{1.82} G^{4,0}_{4,0} \left(1.82 \Biggl\vert \begin{array}{c} \phi,1,1 \\ 0,0,3,
\phi\end{array}\right)
- \\ e^{1.82} G^{4,0}_{4,0} \left(1.82 \Biggl\vert \begin{array}{c} \phi,2,2 \\ 1,1,4,
\phi\end{array}\right)  \Biggl] \approx \\
-0.65 \pi^{-{1\over 2}} m_{e} v_{e}^{3} n_{e} 
\epsilon^{th}  (\beta_{e} \epsilon^{th})^{-1}\\
-0.65 \pi^{-{1\over 2}} m_{e} v_{e}^{3} n_{e}\beta_{e}^{-1}, 
\end{array}
\label{eq:final}
\end{equation}
  where 
\begin{equation}
\begin{array}{c}
G^{m,n}_{p,q}= \left(z \Biggl\vert \begin{array}{c} a_1,.....,a_p \\ b_1,.....,b_q
\end{array}\right)  = \\ ~~~~~~~~~~~~~~~~~~~~~~\frac{1}{2 \pi i}\oint_{L}ds z^{-s} 
\frac{\Pi_{i=1}^{m} \Gamma(b_{i}+s) \Pi_{i=1}^{n} \Gamma(1-a_{i}-s)}
{\Pi_{i=n+1}^{p} \Gamma(a_{i}+s) \Pi_{m+1}^{q} \Gamma(1-b_{i}-s)}
\end{array}
\end{equation}
is the Meijer G function ( Gradshteyn \& Ryzhik 
1980; p. 897, f. 4), and $\phi$ denotes an empty set. 
The numerical evaluation of $G$ was carried out by the use of  Mathematica.

The last equation shows
 the heat flux strongly suppressed compared to the 
Spitzer value when 
the scattering rate by off-axis waves exceeds the electron-ion 
collision  frequency (Levinson \& Eichler 1992).
Previous literature for heat flux inhibiting astrophysical environments
used
a term  $f^{supp}$, the so called suppression factor, 
that multiplied the collision dominated plasma heat flux i.e.
\begin{equation}
q_{\parallel}= -f^{supp} 64 \epsilon^{th}  \pi^{-{1\over 2}} m_{e} v_{e}^{3} n_{e},
\label{eqq:knud}
\end{equation} where we used  eq.~\ref{eq:knud} in eq.~\ref{eq:heat_fon} to 
obtain eq.~\ref{eqq:knud} and the factor $f^{supp}$ was introduced by hand.
The value of $f^{supp}$  obtained from eq.~\ref{eq:final} and \ref{eqq:knud}
is
\begin{equation}
f^{supp}= (64/65)\times 10^{-2}(\epsilon^{th}\beta_{e})^{-1} 
\approx 1 \times 10^{-2}\epsilon^{th}\beta_{e}
\end{equation}
  If the distribution function were not to relax 
to quasilinear marginal stability then there would be less suppression.
However, the non-linear damping terms have been shown to be 
much smaller than  the quasilinear damping term (which in marginal
stability is balanced by a growth term) and thus it can safely 
be argued that the suppression factor computed above is of general
applicability 
(provided that $\beta$ satisfies the required constraints). 
Since $f^{supp}$ has been calculated under the assumption eq.~\ref{eq:symmm}
we suggest the following extrapolation:
\begin{equation}
f^{supp}\approx \frac{1}{1+50(\epsilon^{th}\beta)}. 
\label{supp-1}
\end{equation}
Note that the suppression factor has been calculated relative to a Lorentzian 
plasma. Taking into account that the Spizer-H\"arm heat flux is about 
a factor of five higher relative to a Lorentzian plasma, the suppression factor reads:
\begin{equation}
f^{supp}\approx \frac{1}{1+250(\epsilon^{th}\beta)}. 
\label{supp}
\end{equation}

\section{DISCUSSION AND CONCLUSIONS}\label{sec:con}
Heat-flux inhibition along field lines obtains
when off-axis waves are present in the plasma. 
We have found within quasilinear theory that  whistler 
instability indeed generates  off-axis whistlers.
Thus, heat flux inhibition along field lines by
whistlers should be a common  phenomenon in a weakly 
magnetized 
plasma.  The basic argument for off-axis whistlers is, in qualitative
terms,  that if the 
waves 
were excited only in a narrow cone along the axis, then the 
distribution function would evolve to a state in which the growth
rate of off-axis waves would generally exceed that of the on-axis 
ones.

 One of our assumptions is that the energy density in the magnetic
field of the whistler will not exceed the energy density 
of the background magnetic field. We find that this is generally the case
by a large margin in  tenuous astrophysical plasmas. 
 
While Gary \& Feldman (1977) find 
whistlers inefficient for inhibition of heat flux in the solar wind,
$f^{supp}\sim 1$,
Levinson \& Eichler (1992) find a suppression factor of $10^{-2}
\le f^{supp}\le10^{-1}$, in the interstellar medium (ISM), 
sufficient to explain 
ISM observations. Pistinner et al. (1996) find $1\ge f^{supp}\ge
10^{-4}$ 
in the intracluster medium, ICM, which is enough to resolve many of the
observational problems there. 
In these papers, the level of wave turbulence was calculated 
by assuming a balance between linear growth and non-linear wave removal,
and 
it was noted that the results are suspect when the heat flux so found was
below that needed to drive the whistlers beyond quasilinear saturation.
Here, on the other hand, we have used the criterion of marginal stability
of the least stable whistler wave to derive a relatively simple expression 
for the heat flux  [eq.~\ref{eq:final}, or, equivalently, the
suppression factor 
by which classical heat conduction is suppressed of eq.~\ref{supp}].  The
heat flux that we derive here is at least
that found in our previous papers  inasmuch as   
of the geometry of wave - particle interactions has been considered more 
carefully, i.e. only some particles can interact with some waves, as 
opposed to all particles with all waves via an oversimplified BGK
operator. 
On the other hand,  it will be argued in a subsequent paper that 
the asymmetry of the scattering in the forward and backward 
hemispheres is sufficient to allow for a non-resonant firehose 
instability of the magnetic field.  This causes field line tangling, and 
further inhibits the heat flux. A careful evaluation of the heat flux 
as a function of  plasma parameters will therefore be attempted only 
in the  subsequent paper.

\section{ACKNOWLEDGMENTS} 
We thank A. Levinson for many useful 
discussions about this topic. 
We thank  R. Sunyaev and an anonymous referee for the 
suggestion  to establish
that the energy density stored in waves does not exceed that in 
the total field. Part of this work has been carried out while 
S.P. was a visiting Minerva fellow at the Max-Planck Institute 
f\"ur Astrophysik.

\section{Appendix A}\label{sec:first_appe}
Here we briefly outline the nature of the assumptions 
made in writing the BFPE eq.~\ref{eq:drift_kin}.
The assumption of steady state does not imply that the 
plasma does not evolve with time. Rather it implies that 
any microscopic process occurs on a time scale which is
much faster than the time scale for macroscopic changes 
and thus, temporal macroscopic changes may be ignored to 
first order.

The second assumption implies that energy exchange in a 
collision 
is negligibly small and may therefore be ignored. If the 
electrons
collide only with ions it implies that the ions have infinite mass. 
Thus, electron-electron collisions have been ignored. If the 
plasma is 
collision dominated this assumption modifies the 
collision dominated plasma distribution function by up to a 
factor of five.
A plasma dominated by collective phenomena which leads only 
to pitch 
angle scattering (the case considered in this paper) is 
tantamount 
to the above approximation.

The third assumption is related to the ratio of the gyro-radius 
and the large scale magnetic field variation, the so called 
correlation length (not to be confused with the coherence 
length). 
It implies that this ratio can be effectively set to 
zero.
There is a subtle issue involved. If the plasma is pervaded
by collective magnetic phenomena (electromagnetic waves) 
that have $r_{L}\approx l_{B}$ it seems 
that the approximation which led to eq.~\ref{eq:drift_kin} is 
not valid.
However, this presents a technical problem that can  
be circumvented by a standard procedure 
which allows one to include the small scale magnetic 
fluctuation into
$\nu$ (cf. Melrose 1980). 
Observations suggest that the large scale magnetic field 
variation 
i.e. the magnetic field correlation length, is comparable to the 
electron mean free path.
Thus, we may define:
\begin{equation}
\tilde{\eta}\equiv \frac{r_{L}}{L_{B}}\approx \eta 
\equiv\frac{r_{l}}{\lambda_{e}},
\label{eta}
\end{equation}
where $\lambda_{e}$ is the electron mean free path, and 
$L_{B}$ the magnetic field
correlation length. 
Typical values of $\eta$ are:
\begin{enumerate}
\item $\eta=10^{-12}$~ ISM 
\item $\eta=10^{-16}$ ~ICM.
\end{enumerate}
Assuming that $\tilde{\eta}\approx \eta \rightarrow 0$
one gets the ``adiabatic drift approximation''.

\section{Appendix B}\label{app:A}

Consider the advection of whistlers.
The magnitude of the quasilinear damping term 
$\Omega_{e}\beta_{e}^{-1} \tilde{k} 
\tilde{k}_{\parallel}P(\tilde{k},\chi)$  (eq.~\ref{eq:P_Q})
compared to the advection term in eq.~\ref{eq:waves} is:
\begin{equation}
\frac{ \chi v_g}{2} \frac{\partial \ln (W)}{\partial l_{\parallel}}:
\Omega_{e}\beta_{e}^{-1} \frac{\sqrt{\pi}}{2} 
(1+\chi)^{2} e^{-\frac{1}{(\tilde{k}\chi)^4}}.
\end{equation}
The entire whistler emission process is driven by gradients 
of  the distribution function.
Thus, the shortest length scale for the variation of the 
spectrum is determined by the length scale for the variability of 
the 
distribution function and we have:
\begin{equation}
\begin{array}{c}
\frac{ \chi v_g}{2} \frac{\partial \ln (W)}{\partial l_{\parallel}}
\approx \Omega_{e} \tilde{k}\beta_{e}^{-1} \chi \vert \chi \vert
\left(r_{L} \frac{\partial \ln (W)}{\partial l_{\parallel}}\right)
\approx \\ 
\Omega_{e} \tilde{k}\beta_{e}^{-1} 
\chi \vert \chi \vert \eta \epsilon^{th}. 
\end{array}
\end{equation}
On the axis:
\begin{equation}
\eta \epsilon^{th}\approx 10^{-14}: 1.
\label{eqq:eta_eps}
\end{equation}

\section{Appendix C}\label{app:B}
Levinson (1992) shows that non-linear Landau damping is given by:
\begin{equation}\
\begin{array}{c}
\Gamma^{LD}=\frac{-9 e^{2}}{\sqrt{32 \pi} m_{e} \omega_{p}^{2}}
\int d \chi_{1} d k_{1} \omega_{1}^{w} 
\frac{W(k_{1},\chi_{1}) 
\omega_{2}^{w}}{43 m_{e} v_{e}^{3} \vert k_{\parallel,2}\vert }
e^{-\frac{(\omega_{2}^{w})^{2}}{2\times 1840 ( k_{\parallel,2} v_{e})^{2}}}\\
  \omega^{w}=\omega_{1}^{w}+\omega_{2}^{w}\\
k_{\parallel}= k_{\parallel,1}+ k_{\parallel,2}.
\end{array}
\end{equation} 
Casting the last equation into a non-dimensional form we obtain:
\begin{equation}
\begin{array}{c}
\Gamma^{LD}\approx 
(\Omega_{e}\beta{e}^{-1}) \times \\ 0.14 \beta_{e}^{-2}
\int d \chi_{1}d \tilde{k}_{1} 
\frac{ \tilde{W}(\tilde{k}_{1},\chi)  \tilde{\omega}_{1}^{w} 
\tilde{\omega}_{2}^{w} } 
{\vert  \tilde{k}_{\parallel,2}\vert }
e^{-\frac{1843 \beta_{e}^{-2}(\tilde{\omega}_{2}^{w})^{2} }
{2 ( \tilde{k}_{\parallel,2})^{2} }}\\
\tilde{\omega}^{w}=\frac{\omega^{w}}{\Omega_{e}}.
\end{array}
\end{equation}
The mode coupling term  (Levinson 1992), after it is cast into non dimensional form  reads:
\begin{equation}
\Gamma^{MC}=\frac{1}{16}(\Omega_{e}\beta{e}^{-1})\int d \omega_{1}^{w} 
d \tilde{k}_{1} (1-\chi_{1})^{2}\frac{(\tilde{\omega}_{2}^{w})^{2}}
{\tilde{\omega}_{2}^{w}} \tilde{W}(\tilde{k}_{1},\chi)
\end{equation}
Thus,
\begin{equation}
\Gamma^{w}=
\Gamma^{LD}+\Gamma^{MC}\approx\Gamma^{MC}.
\end{equation}
Levinson (1992)  evaluates the mode coupling term under the assumption that 
the parent waves and the daughter waves have energy of the same order and 
finds:
\begin{equation}
\Gamma^{w}\approx \Gamma^{MC}< (\Omega_{e}\beta{e}^{-1})
\tilde{k}_{\parallel}^{2} \frac{\tilde{\nu}^{w,s}}{\tilde{\nu}^{coll}}\eta.
\end{equation}
Consider now the ratio of the quasilinear damping term to the non-linear 
damping term.
\begin{equation}
\begin{array}{c}
(\Omega_{e}\beta{e}^{-1}) : (\Omega_{e}\beta{e}^{-1})
 \frac{\tilde{\nu}^{w,s}}{\tilde{\nu}^{coll}}\eta \\
1: \frac{\tilde{\nu}^{w,s}}{\tilde{\nu}^{coll}}\eta 
 \approx 10^{-8}\ll 1
\end{array}
\end{equation}

\section{Appendix D}\label{app:C}
Here we consider how the exact numerical value of 
$\frac{Y^{2}}{Y^{1}}$ is obtained. Although in principle we deal with an 
integral equation, it can be reduced by the procedure outlined below.
Let 
\begin{equation}
a=2-\frac{Y^{2}}{Y^{1}},
\end{equation}
then substituting, eqs~\ref{eqq:sol_s}
and~\ref{eqq:sol_a} into eq.~\ref{eq:cur_ints} we find,
\begin{equation}
\begin{array}{c}
Z(\tilde{v})= (2 \epsilon^{th} \beta_{e})^{-1}\frac{14}{51}
\tilde{v}^{-3} \\  \Biggl[a \left(1+\frac{\tilde{v}^{2}}{a}\right)\log 
\left(1+ \frac{\tilde{v}^{2}}{a}\right)- \tilde{v}^{2}\Biggl].
\end{array}
\end{equation}
One can now proceed and calculate formally the value of $Y^{2} $ and
$Y^{1}$.
Substituting the expression for $Z(\tilde{v})$ 
into eq.~\ref{eq:cur_ints}  one 
finds after some algebra
\begin{equation}
\begin{array}{c}
 \frac{51}{7} \epsilon^{th} \beta_{e} Y^{2}=\\ 
1-0.25[a+a^{2}+2e^{a}
G^{4,0}_{4,0} \left(a \Biggl\vert \begin{array}{c} \phi,1,1 \\ 0,0,3,
\phi\end{array}\right)-\\
e^{a} G^{4,0}_{4,0} \left(a \Biggl\vert \begin{array}{c} \phi,1,1 \\ 0,0,2,
\phi\end{array}\right)]\\
 \frac{51}{7} \epsilon^{th} \beta_{e} 
Y^{1}=0.5[1-e^{a} G^{4,0}_{4,0} \left(a \Biggl\vert \begin{array}{c} \phi,1,1 \\ 0,0,2,
\phi\end{array}\right)]
\end{array}
\end{equation}
Subtracting from $2$ the formal ratio of $Y^{2}/Y^{1}$ one obtains
a transcendental equation for $a$. This equation reads:
\begin{equation}
\begin{array}{c}
a-2+2\Biggl[1-0.25(a+a^{2}+2e^{a} G^{4,0}_{4,0} \left(a \Biggl\vert \begin{array}{c} \phi,1,1 \\ 0,0,3,
\phi\end{array}\right)-\\ 
e^{a} G^{4,0}_{4,0} \left(a \Biggl\vert \begin{array}{c} \phi,1,1 \\ 0,0,2,
\phi\end{array}\right) 
\Biggl] \\
\Biggl[1-e^{a}
G^{4,0}_{4,0} \left(a \Biggl\vert \begin{array}{c} \phi,1,1 \\ 0,0,2,
\phi\end{array}\right)
\Biggl]^{-1} =0.
\end{array}
\end{equation}
Solving this equation numericaly, we find $a=1.82$.

\bsp % ``This paper has been produced using the ...''


\begin{thebibliography}{}

\bibitem{}
Blandford, R. D., \& Eichler D., 1987,  RMPh, 1987, 154, 2
\bibitem{}
Balbus, S. A. \& McKee, C.F., 1982, ApJ, 252, 529
\bibitem{}
Bandiera, R. \& Chen, Y., 1994a, A\&A, 284, 629
\bibitem{}
Bandiera, R. \& Chen, Y., 1994b, A\&A, 284, 637
\bibitem{}
Begelman, M, C. \& McKee, C. F., 1990, 358, 375
\bibitem{}
Binney, J. \& Cowie, L.L., 1981, ApJ, 247, 464
\bibitem{}
Cowie, L.L. \& Binney, J., ApJ, 1977, 215, 723
\bibitem{}
Cowie, L.L. \& McKee, C.F., 1977, ApJ, 211, 135
\bibitem{}
Cowie, L.L. \&  Songaila, A., 1977, Nature,  266, 510
\bibitem{}
Chun, E. \& Rosner, R. 1993, ApJ, 408, 678
\bibitem{}
Canizares, C. R. Markert, T.H. \& Donahue, M.E.,  1988, In 
cooling 
flows in clusters of galaxies. pp. 63-72, (ed: Fabian)
\bibitem{}
Dalton, W. W., \& Balbus, S. A., 1993, ApJ, 404, 625.
\bibitem{}
Fabian, A.C., \& Nulsen, P.G.E., 1977, MNRAS, 180, 479
\bibitem{}
Fabian, A.C., Nulsen, P.E.J \& Canizares, C. R., 1984, Nature, 
180, 733
\bibitem{}
Fabian, A.C., Nulsen, P.E.J \& Canizares, C. R., 1991, A\&AR, 
2, 191
\bibitem{}
Fabian, A. C., 1994, ARA\&A, 32, 277
\bibitem{}
Field, G. B., 1965, ApJ, 142, 531
\bibitem{}
Fukazawa, Y., 1994, PASJ, 46, 55
\bibitem{}
Galeev, A. A., \& Natanzon, A.M., 1984, Dphy, 275, 6.
\bibitem{}
Goldstein, M. L., Klimas, A. J. \& Sandri, G., 1975, ApJ
\bibitem{}
Gary, S.P. \& Feldman, W.C., 1977, JGR, 82, 1087
\bibitem{}
Gary, S.P., Schime, E.E., Phillips, J.L.,  
\bibitem{}
Gradshteyn, I.S., \& Ryzhik, I.M., 1980, (Academic:New-York)
\bibitem{}
Gray, D. R. \& Killkenny, J. D., 1980, Plas. Phys., 22, 81. 
Feldman, W.C., 1994, JGR, 99,23391 
\bibitem{}
Jafelice, L. C.,  1992, AJ, 104, 1279
\bibitem{}
Levinson, A. \& Eichler, D., 1992, ApJ, 387, 212
\bibitem{}
Levinson, A., 1992, ApJ, 401, 73
\bibitem{}
Loewenstein, M. \& Fabian, A.C., 1990, MNRAS, 242, 120
\bibitem{}
Melrose D.B., 1980, Plasma Astrophysics,  (New-York: Gordon)
\bibitem{}
McKee, C. F. \& Begelman, M, C., ApJ, 358, 392
\bibitem{}
McKee, C.F. \& Cowie, L.L., 1977, ApJ, 215, 213
\bibitem{}
McKee, C.F. \& Ostriker,  J.P., ApJ, 218, 148
\bibitem{}
Montgomery, M. D. Bame, S. J. \& Hundhausen, 1968, JGR, 73,
4999
\bibitem{}
Pistinner, S. L.  \& Shaviv,  G., 1996, ApJ, 459, 147
\bibitem{}
Pistinner, S. L., Levinson, A.  \& Eichler, D., 1996, ApJ, 467, 
162
\bibitem{}
Pistinner, S. L.\& Eichler, D., 1998, Preprint
\bibitem{}
Rosner, R. \& Tucker, W. H., 1989, ApJ, 338, 761
\bibitem{}
Sagdeev, R.Z., \& Galeev, A. A., 1968, Nonlinear Plasma 
\bibitem{}
Schime, E.E, Bame, S.J.  Feldman, W.C., 
Gary, S.P., Phillips, J.L., 1994, JGR, 99, 23401 
Theory (New-York: Benjamin, inc.) 
\bibitem{}
Slavin, J., D., \& Cox, D., P., 1992, ApJ, 392, 131.
\bibitem{}
Spitzer, L. Jr., 1962, Physics of Fully Ionized Gases (New-York: 
Wiley)
\bibitem{}
Stix, T. H., 1992, Waves in Plasmas, (AIP: New-York).
\bibitem{}
Tao, L, 1995, MNRAS, 275, 965
\end{thebibliography}
\end{document}
\end